\documentstyle[11pt]{article}
\newcommand{\infig}[2]{\begin{center}\mbox{\epsfxsize #2
\epsfbox{#1}}\end{center}}

\input {epsf}

\newcommand{\beq}{\begin{equation}}
\newcommand{\eeq}{\end{equation}}
\newcommand{\beqa}{\begin{eqnarray}}
\newcommand{\eeqa}{\end{eqnarray}}

\def\opone{\leavevmode\hbox{\small1\kern-3.8pt\normalsize1}}
\def\text{ }
\def\textit{\it }
\def\textbf{\bf }

\input{tcilatex}

\begin{document}

\title{Experimental test of non-local quantum correlation in relativistic
configurations}
\author{H. Zbinden, J. Brendel, N. Gisin, W. Tittel \\
%EndAName
Group of Applied Physics, University of Geneva}
\maketitle

\begin{abstract}
We report on a new kind of experimental investigations of the tension
between quantum nonlocally and relativity. Entangled photons are sent via an
optical fiber network to two villages near Geneva, separated by more than 10
km where they are analyzed by interferometers. The photon pair source is set
as precisely as possible in the center so that the two photons arrive at the
detectors within a time interval of less than 5 ps (corresponding to a path
length difference of less than 1 mm ). One detector is set in motion so that
both detectors, each in its own inertial reference frame, are first to do
the measurement! The data always reproduces the quantum correlations, making
it thus more difficult to consider the projection postulate as a compact
description of real collapses of the wave-function.
\end{abstract}

\section{Introduction}

%=====================
Since the famous article by Einstein, Podolsky and Rosen (EPR) in 1935 \cite
{EPR35} ''quantum non-locality'', received quite a lot of attention. They
described a situation where two entangled quantum systems are measured at a
distance. The correlation between the data can't be explained by local
variables, as demonstrated by Bell in 1964 \cite{Bell64}, although they
can't be used to communicate information faster than light. Indeed, the data
on both sides, although highly correlated, is random. There is thus no
obvious conflict with special relativity. In a realist's view the
correlation could be ''explained'' by an action of the first measurement on
the second one. This explanation is intuitive, but, if the distance between
the two detection events is space-like, incompatible with relativity.

This tension between quantum mechanics and relativity has long been studied
by theorists, see among many others \cite
{Shimony83,AharonovAlbert81,Hardy92,IP99,ghirardietal1990}. But it clearly
deserves to be analyzed experimentally and the first realization of a new
kind of test is the main result of this article. Let us recall that the
tension is between the collapse of the quantum state (also called
''objectification'' \cite{Busch}, or ''actualization of potential
properties'' \cite{Jauch,Piron,Shimony}) and Lorentz invariance: since the
collapse instantaneously affects the state of systems composed of distance
parts, it can't be described in a covariant way. From this observation one
may conclude either that there is no collapse or that this tension indicates
a place to look for new physics. The assumption that there is no real
collapse sounds strange to us, though, admittedly, it deserves to be
developed \cite{deutsch}. Anyway, in this article we follow the intuition
that the tension between quantum mechanics and relativity is a guide for new
physics. In order to test this tension, we design an experiment in which
both quantum nonlocality and relativity play a crucial role. By quantum
nonlocality we mean here measurements on two systems that are spatially
separated, but still described as one global quantum object characterized by
one state vector (i.e. the two systems are entangled). For relativity we
like a role more prominent than mere spatial separation, as in tests of Bell
inequalities. Central to relativity (at least to special relativity) is the
relativity of time, which may differ between inertial frames with relative
velocities. Hence, we like to have the observers of the two quantum systems
not only space-like separated, but moreover in relative motion such that the
chronology of the measurement events is relative to the observer: each
observer in his own inertial frame performs his measurement before the other
observer. In such a situation the concept of ''collapse'' is even weirder:
both observers equivalently claim to trigger the collapse first! Actually
one can then argue, following Suarez and Scarani \cite{Suarez97}, that in
such a situation each measurement is independent from the other and that the
outcomes are uncorrelated (more precisely, that only classical correlation
remain, the quantum correlation carried by the wave-function being broken,
see appendix A).

Let us elaborate on the intuition that motivates our experiment. Each
reference frame determines a time ordering. Hence, in each reference frame
one measurement takes place before the other and can be considered as the
trigger (the cause) of the collapse. The picture is then the following: the
first measurement produces a random outcome with probabilities determined by
the local quantum state (the entire state is not needed to compute the
probabilities, the local state obtained by tracing over the distant system
suffices). When the outcome is produced, the global quantum state is
reduced. This is the controversial part of the process since this reduction
happens instantaneously in theory and faster than light according to
experimental tests of Bell inequality. The second measurement can then be
described like the first one: it produces a random outcome with
probabilities determined by the local state. The only difference with
respect to the first measurement is that the local state of the second
measurement contains ''information'' about the outcome of the first
measurement. Since this ''information'' can not be used to transmit a
classical message, there is no direct conflict with relativity and we term
this ''information'' as quantum information. Since in our experiment the
time intervall between the two meaurements is minimized, it will supply a
lower limit of the speed of this quantum information. So far the described
picture of the collapse is compatible with relativity, and with all
experiments performed so far. But let us now assume that the two observers
are in relative motion such that they disagree on the time ordering of the
measurements. Then, it is plausible that each measurement produces random
outcomes with probabilities determined by the local sates (see appendix A): 
\begin{eqnarray}
Prob(++|before-before) &=&<P_{\alpha }\otimes \leavevmode\hbox{\small1%
\kern-3.8pt\normalsize1}>_{\psi }<\leavevmode\hbox{\small1\kern-3.8pt%
\normalsize1}\otimes P_{\beta }>_{\psi } \\
&=&Tr(P_{\alpha }\rho _{1})Tr(P_{\beta }\rho _{2})
\end{eqnarray}
where $\rho _{1}=Tr_{2}(\psi \psi ^{\dagger })$ and $\rho _{2}=Tr_{1}(\psi
\psi ^{\dagger })$ are the local states obtained by tracing over the other
sub-system and $P_{\alpha }$ and $P_{\beta }$ the projectors corresponding
to the two measurements. For singlet states, for example, the probability of
(++) outcomes would be $\frac{1}{4}$, independently of the orientations of
the analyzers. The above prediction, based on plausible intuition inspired
by the work of Suarez and Scarani \cite{Suarez97}, differs from the quantum
predictions. However, it does not contradict any experimental result and
deserves thus to be tested experimentally.

At first sight the experiment may seem exceedingly difficult. First, because
it requires relativistic speeds in order to have different time ordering.
Next, because it requires the precise characterization of the vague concepts
of "observer" and "measurement". In the next section, we argue that the
first difficulty can be mastered and that the second one offers the
possibility to elaborate on these delicate concepts in an experimentally
defined context (much more promising than mere theoretical elecubrations).
Then, sections 3 and 4 describe the technical aspects of the experiment,
while section 5 contains the measurement results.

\section{A Bell experiment with a moving observer}

%===================================================
Consider the following situation depicted in Fig. 1: A source S emits two
photons that travel to the analyzers at Alice and Bob respectively. Bob is
moving away from Alice with the speed $v$. Let the event A, the detection of
the first photon by Alice, be at the origin of reference frames of Alice and
Bob ($A_{a}=(x=0,t=0),A_{b}=(x`=0,t`=0)$ ). \ For the event B, the detection
of the second photon by Bob, we note $B_{a}=(x=L,t=\delta t)$, where L is
the distance between Alice and Bob at time t=0 and $\delta t$ is the (small)
time difference between B and A in the reference frame of Alice. The time
difference between B and A in the reference frame of Bob, is

\begin{equation}
\delta t^{\prime }=\frac{\delta t-\frac{vL}{c^{2}}}{\sqrt{1-\frac{v^{2}}{%
c^{2}}}}\cong \delta t-\frac{vL}{c^{2}}
\end{equation}

Hence, if

\begin{equation}
\frac{vL}{c^{2}}\geq \delta t\geq 0  \label{moving}
\end{equation}
then the time ordering of A and B is not the same in the reference frames of
Alice and of Bob. As a consequence both observers have the impression to
perform the measurement first, hence to provoke the collapse of the
wavefunction first. And one can argue, as we have done in the introduction,
that in such a situation the non-local correlation should disappear.

Reasonably achievable speeds $v$ are in the order of 100m/s. This means that
if you intend to perform the experiment in the lab with separations L say
20m, $\delta t $ must be smaller than 20 fs, corresponding to a distance in
air of 6 $\mu $m. Such a short distance or time difference is only useful if
the photon is localized as well as that. A coherence length of 6$\mu $m
demands a bandwidth of 150 nm, which is hardly achievable for photon pairs.
In conclusion, one should go to larger distances L which requires optical
fibers. For L=10km, $\delta t$ must be shorter than 10 ps or 2mm of optical
fibre. So you need two optical fibre links of at least 5 km (installed fiber
are not straight lines!) that are equal to about 1 mm in length. At the same
time you have to make sure that the photons are not delocalized due to the
chromatic dispersion. The next section describes how we can achieve that.

Before, we have to discuss the question {\it ''what is an observer?''}, and 
{\it ''where does a collapse of the wavefunction take place?''}. Is it in
the physicist's mind, in the photon counter, or already at the beamsplitter
or polarizer? The answer to such questions is usually not of practical
relevance; however, for our kind of experiment the answer is crucial. It
determines which part of the experiment should be moving and to which point
the optical paths must have the same length. In this article, we assume that
the effect occurs at the detector, when the irreversible transition from
''quantum'' to ''classical'' occurs. So we will have a situation like in
Fig. 2a. Alice and Bob each have two detectors ( $A_{+},A_{-},B_{+},B_{-}$),
two of which ($A_{+,}B_{+}$) are at precisely the same distance from the
source. There are four possible outcomes with the corresponding
probabilities. Normalization (conservation of particle number) imposes: 
\begin{eqnarray}
P_{A_{+},B_{+}}+P_{A_{+},B_{-}}+P_{A_{-},B_{+}}+P_{A_{-},B_{-}} &=&1
\label{norm} \\
P_{A_{+}}+P_{A_{-}} &=&1 \\
P_{B_{+}}+P_{B_{-}} &=&1
\end{eqnarray}
The last two eqs. imply that if the photon is not detected by $A_{+}$, then
it is necessarily detected by $A_{-}$ (assuming perfectly efficient
detectors), and similarly for Bob's detectors. Moreover, for maximally
entangled states, as in our experiment, all 4 detectors have a probability $%
\frac{1}{2}$ of detecting a photon, independently of the interferometers
settings: 
\begin{equation}
P_{A_{+}}=P_{A_{-}}=P_{B_{+}}=P_{B_{-}}=\frac{1}{2}  \label{Pdemi}
\end{equation}
With all detectors at rest, we can expect correlation of the events as a
function of phase in the interferometers (or angles of the polarizers), as
demonstrated in our previous experiment \cite{Tittel2,Tittel3}: 
\begin{equation}
P_{A_{+},B_{+}}^{QM}=P_{A_{-},B_{-}}^{QM}=\frac{1+\cos (\delta _{a}+\delta
_{b})}{2}\ \ {and\ }P_{A_{+},B_{-}}^{QM}=P_{A_{-},B_{+}}^{QM}=\frac{1-\cos
(\delta _{a}+\delta _{b})}{2}
\end{equation}
where the suffix $QM$ stands for Quantum Mechanics. If now, the detector $%
B_{+}$ is moving with respect to $A_{+}$ one might argue that correlations
disappear (see the introduction and appendix A), i.e. that 
\begin{equation}
P_{A_{+},B_{+}}^{bb}=const.=\frac{1}{4}  \label{quart}
\end{equation}
where the suffix $bb$ stands for before-before. Using (\ref{Pdemi}) one has $%
P_{A_{+},B_{-}}=const.=\frac{1}{4}$ and, with help of (\ref{norm}) one
deduces $P_{A_{-},B_{-}}=const.=\frac{1}{4}$. Consequently, one can test the
prediction (\ref{quart}) by only looking at the coincidences between $A_{-}$
and $B_{-}$, the two detectors at rest! These detectors just have to be
further away from the source than $A_{+}$ and $B_{+}$ to assure that the
collapse occurs at $A_{+}$ or $B_{+}$.

The next question reads {\it What is a detector?}. One definition could be:
it is any physical system in which the photon is irreversibly absorbed and
transformed into a classical signal. The essential part is the irreversible
process, the absorption of the photon in a solid and the transformation of
its energy into heat. Since, as we have seen above, we don't need to read
the signal of the moving detector, the detector can be turned off or only
consist of a non-fluorescent, black paint absorber (one can imagine that we
could measure the temperature increase due to the absorption of the photon).
Accordingly, detectors $A_{+}$ and $B_{+}$ can be just absorbing black
surfaces The black surfaces provoke the collapse, even in the case that the
photons go to the detectors! This considerably facilitates the experiment:
First, we don't need to identify precisely the absorbing layer of the
photodiode. Second, the moving detector will be a spinning wheel. The rim of
the wheel is painted black and the end of the optical fiber is pointed from
outside on it (see Fig. 2b). Note that during the 50 $\mu $s time of flight
of the photon from the source, the wheel's edge moves by about 5 mm, thus
our ''detector'', the molecules of the black paint, make in a good
approximation a linear movement, defining the inertial reference frame. We
don't need electrical or optical contact, neither cooling of our spinning
wheel as if a real detector was mounted.

We admit that the above argument is questionable and that it would be nicer
to have real, moving photon counters that make ''click''. However, the
argument is fair and it clearly turns a nightmare experiment into a feasible
one. Let us also note that in their original work, Suarez and Scarani
proposed, inspired by Bohm's model, that the relevant device is the
beamsplitter and not the detector. Thus, our\ experiment is not a test of
their model.

\section{\protect\bigskip Equalizing two fiber optical links}

In order to perform the experiment, the source has to be set precisely at
the center so that in the Geneva reference frames both photons are analyzed
within a few ps, corresponding to about a millimeter over a fiber length of
more then 18 km. In this section we describe how we achieved such a
precision. Clearly, the chromatic dispersion which spread the photon
wave-packet is also a serious concern.

We used installed telecom fibres and work with photons at wavelengths around
1300 nm. The relative group delay $\tau (\lambda )$ of a pulse, expressed in
[ps/km], can be well fitted by the Sellmeier equation: 
\begin{equation}
\tau (\lambda )=a\lambda ^{2}+b\lambda ^{-2}+c  \label{tau1}
\end{equation}

The dispersion coefficient $D$ is defined as $D=\frac{d\tau }{d\lambda }$
has the units $ps/nm\cdot km$ and goes to zero at the zero dispersion
wavelength $\lambda _{0}$. The parameter $S_{0}$ is the slope of the
dispersion at $\lambda _{0}.$ For standard telecom fibers $\lambda _{0}$ is
situated around 1310 nm and S$_{0}$ is close to 0.08 ps/km nm$^{2}.$
Equation (\ref{tau1}) can be rewritten in terms of $S_{0}=8a$ and $\lambda
_{0}=\sqrt[4]{b/a}$: 
\begin{equation}
\tau (\lambda )=\tau _{0}+\frac{S_{0}}{8}(\lambda -\frac{\lambda _{0}^{2}}{%
\lambda })^{2}\approx \tau _{0}+\frac{S_{0}}{2}(\lambda -\lambda _{0})^{2}
\label{disp}
\end{equation}
The first term (the group velocity delay) can be adjusted with a precison
around 100 $\mu m$ (see below): 
\begin{equation}
\tau (\lambda _{0}^{A})l^{A}\approx \tau (\lambda _{0}^{B})l^{B}
\end{equation}
where $l^{A}$, $l^{B}$, $\lambda _{0}^{A}$ and $\lambda _{0}^{B}$ denote the
lengths and the zero dispersion wavelengths of the fibres going to Alice and
Bob, respectively. However, a simple estimate of the second term (chromatic
dispersion) shows that a photon centered at $\lambda _{0}$ with a bandwidth
of say 50 nm would suffer a spread of 245 ps per 10 km, which is much more
than the maximum 10 ps target. Fortunately, working with photon pairs the
major part of dispersion can be cancelled due to the energy correlation of
the photons. Since $\omega _{p}=\omega _{s}+\omega _{i}$ the delays
undergone by the signal and idler photons can be equalized: 
\begin{equation}
S_{0}^{A}(\lambda ^{A}-\lambda _{0}^{A})^{2}\approx S_{0}^{B}(\lambda
^{B}-\lambda _{0}^{B})^{2}
\end{equation}
For this we chose fibers such that $\lambda _{0}^{A}\approx \lambda _{0}^{B}$%
, $S_{0}^{A}\approx S_{0}^{B}$, and tuned the pumpwavelength to obtain $%
2\lambda _{p}=\frac{\lambda _{0}^{A}+\lambda _{0}^{B}}{2}$ (note, that eq. (%
\ref{disp}) is in function of $\lambda $ whereas the signal and idler
photons are symmetric in $\omega $, but makes only a second order difference
). The difference in the group delay is then:

\begin{equation}
\delta t=l^{A}(\tau _{0}^{A}+\frac{S_{0}^{A}}{2}(\lambda ^{A}-\lambda
_{0}^{A})^{2})-l^{B}(\tau _{0}^{B}+\frac{S_{0}^{B}}{2}(\lambda ^{B}-\lambda
_{0}^{B})^{2})\approx 0
\end{equation}

Accordingly, in theory the quadratic term of (\ref{disp}) cancels and higher
order terms are limiting. In practice, however, the precision with which one
can measure and equalize the $\lambda _{0}$'s of the fibers determine $%
\delta t$. For a given spectral distribution of the photon pairs, determined
by the transmission curve of an interference filter a corresponding
distribution of the $\delta t$ in the group delay (see Fig. 3 ) is obtained.
Table 1 gives two figures of merit for the temporal spread in ps per km for
different deviations of center wavelength of the photon pairs from the mean\
zero dispersion wavelength ($2\lambda _{p}-\frac{\lambda _{0}^{A}+\lambda
_{0}^{B}}{2}$) and bandwidths $\Delta \lambda $ (FWHM) of the filter. These
are the $\Delta \tau _{\max }$ , the maximum difference of $\delta t$ \ for
95\% of the photons that are within the $2\lambda _{p}\pm 2\sigma $
intervall and $\Delta \tau =2\sqrt{\int (\delta t(\lambda )-\overline{\delta
t})^{2}p(\lambda )}$, the mean square deviation. This figures depend
essentially on $\Delta \lambda $ and and in a first approximation not on $%
\lambda _{0}^{A}-\lambda _{0}^{B}$.

\bigskip

\begin{tabular}{||l|l|l|l||}
\hline\hline
$2\lambda _{p}$ & $\Delta \lambda =$ 10 nm & $\Delta \lambda =$ 40 nm & $%
\Delta \lambda =$ 70 nm \\ \hline
1309.0 & 0.95 (1.90) ps & 3.3 (6.0) ps & 4.4 (6.3) ps \\ \hline
1309.5 & 0.47 (0.93) ps & 1.4 (2.2) ps & 2.9 (2.6) ps \\ \hline
1310.0 & 0.013 (0.026) ps & 0.8 (1.7) ps & 6.7 (13.6) ps \\ \hline
1310.5 & 0.49 (0.99) ps & 2.6 (5.5) ps & 7.4 (15.8) ps \\ \hline
1311.0 & 0.97 (1.94) ps & 4.5 (9.4) ps & 0.95 (1.90) ps \\ \hline\hline
\end{tabular}
\newline

\bigskip Table 1: $\Delta \tau $ ($\Delta \tau _{\max })$ per km of fiber
for different $\lambda _{p}$ and $\Delta \lambda $ (FWHM). $\lambda
_{0}^{A}=\lambda _{0}^{B}=1310$ $nm.$

The chromatic dispersion was first measured with a commercial dispersion
measurement apparatus (Anritsu ME 9301A) . Unfortunately, this apparatus
revealed variations of the zero-dispersion wavelength of up to 2 nm. Next,
we built up our own apparatus using a Delay generator (Standford 530), a
pulsed LED, a tunable filter(JDS), an actively gated InGaAs APD and a time
to amplitude converter (TAC, Tenelec TC836) \cite{cd-proto}. We obtained a
reproducibility of 0.1 nm and estimated the absolute error as 0.2 nm \cite
{BrendelOFMC99}. The length measurement was done in a first step with a
home-made OTDR setup similar to that used for the dispersion measurement. It
achieved a precision of 1-2 mm. In a second step we used a low coherence
reflectometer (an interferometer with a scanning mirror) to determine the
path difference to a precision of about 0.1 mm (Figure 4 shows a typical
scan) \footnote{%
Since the absorbers don't reflect the light, the measurement has to be
performed in two steps: For both interferometers, we measured the path
lengths from the input of the circulator to the absorbing surfaces\ with a
precision of about 50 $\mu $m. In addition we determined the length of two
pigtails with mirrored fiber ends with the same precision. This allowed us
then to measure the path length difference between link A and B by replacing
the interferometer by the two calibrated pigtails with mirrors.}. The
standard resolution of about 20 $\mu m$ could not be obtained, since we had
to reduce the spectral width of the LED with a tunable filter of 2 nm width
(FWHM) in order to see interference. If the dispersion properties of two
fibers were perfectly identical, the wavelength of the LED wouldn't matter.
We limited a possible shift of group delay by centering the filter at 2$%
\lambda _{p}$. We estimated the error of $2\lambda _{p}$ as 0.2 nm and we
conservatively assumed that $2\lambda _{p}-\frac{\lambda _{0}^{A}+\lambda
_{0}^{B}}{2}\approx 0.5$ $nm.$ For relative high differences $\lambda
_{0}^{A}-\lambda _{0}^{B}\leq 1$ nm we then obtained a maximal shift of 0.1
ps/km what is negligable. Concerning the temporal spread of the photons we
obtained according to Table 1 a spread $\Delta \tau $ of about 5 ps (for 10
km of fiber) if the bandwith of the downconverted photons is limited to 10
nm.

We performed our experiments between three Swisscom stations in Geneva and
Bernex and Bellevue separated by 10.6 km bee-line. We obtained 9.53 km for
link A (Geneva-Bernex) and 8.23 km for the link B (Geneva-Bellevue). We
added 500 m dispersion shifted fiber to link A and 1.80 km of standard fiber
to link B equalize roughly the length and as precisely as possible $\lambda
_{0}$ and ended up with $\lambda _{0}^{A}=1313.0$ nm and $\lambda
_{0}^{B}=1313.3$ nm. Two meters of fiber were mounted on a rail in order to
adjust the length of link A by pulling or releasing the fiber. In Bellevue,
the distance between the end of the fiber and the black wheel could also be
adjusted within a range of a few mm.

\bigskip

\section{Experimental setup}

%===========================
The experimental setup was similar to our Franson-type Bell-experiment
presented earlier in more detail \cite{Tittel2}, \cite{Tittel3}, see Fig. 5.
The parametric downconversion source consisted essentially of a 655 nm diode
laser (30mW Mitsubishi) with external grating and a KNbO$_{3}$ crystal
(length 10 mm, cut at $\theta =33^{o}$). The analyzers were two Michelson
interferometers with Faraday mirrors. We used optical circulators at the
input ports in order to access to both outputs of the interferometers. At
the circulator output ports we had our absorbers, a black scotch tape at A,
the black wheel at B. These two surfaces were at exactly the same distance
from the source. At the other output ports we connected our photon counters
(passively quenched NEC Ge APD's), making sure that they were further away
from the source than the absorbers. Any detection triggered a laser pulse
that was sent back to the source through another optical fibre. A TAC
(Tenelec TC863) with Single Channel Analyzer selected the events with the
right time interval, corresponding to two interfering possibilities when the
photons take either both the short or both the long arm of the
interferometers. We obtained typically 20 kcts/s single count rate plus 45
kcts/s dark count rates. This lead to a mean value of 10 coincidences per
second. With the 10 nm FWHM interference filter inserted, we obtained 2 kcts
singles and about 3 coincidences per second. The interferometers were
temperature controlled. Interferometer A was kept at a constant temperature
of 30$^{o}$C. The temperature of interferometer B scanned between 30.5 and
37.5 $^{0}$C. This produced a variation of the phase of about 10$\cdot 2\pi $
and therefore allowed us to record the coincidences as a function of the
phase. The path length difference was measured to be equal when the
temperature was $34^{o}$C. The wheel was a 20 cm diameter aluminium disk of
1 cm thickness directly driven by a brushless 250W DC motor (Maxon EC). It
turned vertically at 10000 rpm leading to a tangent speed of 105m/s. The
fiber pointed from the top on the blackened outer rim of the wheel. The
wheel placed at Bellevue was oriented with a compass to make it run away or
towards the other observer at Bernex.

\section{Measurements and Results}

We measured the path length difference between link A and B with the low
coherence reflectometer. We found that the measurements were quite
reproducible on short term, however, the length difference could vary by up
to a few mm per hour. Actually we found that Bernex was drifting further
away during the daytime, probably due to the fact that link A was more
exposed to the daily temperature rise.

One possibility to test for the breakdown of the quantum correlation would
be to measure the 2-photon interference visibility, move one absorber
slightly closer, repeat the measurement and so on. As discussed above ( see
Table 1), to limit dispersion effects and to obtain a good timing, we
introduced a 10 nm (FWHM) interference bandpass filter after the source,
reducing the coincidence count rate to some 3 cts/s. Hence to get a
reasonable measurement statistics we needed some 100 sec integration time
per measurement point, and some 10 points to see one interference fringe to
determine the visibility, hence the measurement time was about 20 minutes.
Since we couldn't simultaneously measure the path difference and since the
uncertainty in distance after 20 min was more than 1 mm, it was difficult to
make a scan in distance with a spatial resolution better than 1mm. So we
decided to renounce to a manual distance scan and to take profit of the
natural temperature induced drift. This drift proved to be monotonous in one
sense during the day and in the other sense during the night. So we almost
aligned the paths knowing that due to daily the drift will perfectly aligned
in certain moment later in time. We started then to record the interference
fringes of the coincidences by homogeneously varying the phase. Finally we
confirmed with a second position measurement after a few hours that the path
lengths really passed through the presumed equilibrium. Then, we analyzed
the interference fringes and looked for periods of reduced visibilities
during the measurement.

Figure 6 shows typical data taken over 6 hours while the optical link to
Bernex lengthened by 2 mm with respect to the one to Bellevue. The
difference of the optical path lenghts, expressed in $\delta t$, was varying
from + 8 to -1.3 ps. Positive values mean that the detections occured first
in Bernex. In the moving Bellevue reference frame the detections happened
first in Bellevue over the entire scan range, as indicated by the negative
values of $\delta t^{\prime }$on the upper time scale. Despite this
different time ordering no reduced visibility is observed. Inevitably, the
curves show high statistical fluctuations due to the low count rate. In
spite of this, one can state that the visibilitiy of the two photon
interferogram remains constant. Especially, a reduced visibility over a scan
span of 1 mm (corresponding to 5 ps) should easily be noticed. After
substraction of the 237$\pm 5$ cts/100s accidental coincidences, the fit of
Fig 2 shows a\ constant fringe visibility of 83\%, large enough for a
violation of Bell's inequality. Note however, that hidden variables are no
issue in this work.

Figure 7 shows a scan over a longer period (14 hours overnight) and larger
scan range. This time the wheel was turning in the other direction (towards
Bernex) and the optical link to Bernex shortened by 20mm. Comparing the two
time scales we find a period in the left part of the scan where we have
negative values below and positive ones above. Hence, we have again a
different time ordering, now both observers suppose to make the measurement
after the other. Again, there is no evident drop of the visibility over a
period of 10 ps, i.e. over roughly two fringes, assuming a rather homogenous
scan rate. This scan over such a long period shows the problems of the
experiment. The period of the fringes can vary due to small drifts of the
pump laser frequency. The photon counters may have varying efficiency and
dark count level. The slowly mounting line of accidentals is due to the fact
that we measured higher darkcount rates at the end of the experiment,
possibly due to increased temperature of one of the detector, since there
was little liquid nitrogen left. Finally, the visibility may slightly\
decrease due a drift of the coincidence window discriminating the three
temporal peaks of the Franson interferogram. Moreover, during a long scan
some artefact are probable, as the the considerably higher peak after one
third of the scan (sombody switching the lights on in the Swissom station?).
The arrow indicates the moment when the scan of the phase by scanning the
temperature of the Bernex interferometer changed the direction. Altogether
we recorded over 20 tracks similar to those presented in Fig. 6 and 7 and no
reproducable effect on the visibility could be determined.

The Figures 6 and 7 can also be used to estimate the lower bound for the
speed of quantum information. At a certain time the two paths are perfectely
equal and the lower bound could be arbitrarly high. In practice two factors
limit this lower bound. First, we assume that at least one fringe should
vanish to be able to state a reduced visibility. In Fig. 6, for instance, we
observe 7 fringes for 10 ps delay. So the minimum time difference is say 1.5
ps. The second factor, the temporal spread of the photons, is the
determining one in our case. We estimate it to be smaller than 5 ps. The
lower speed limit becomes then:\ 
\begin{equation}
\frac{10.6\ km}{\ 5\ ps}\approx 2\cdot 10^{15}\frac{m}{s}=\frac{2}{3}10^{7}c
\end{equation}
We also performed measurements with two Ge APD's precisely aligned instead
of the absorbing surfaces showing the same evidence. Further we removed the
interference filter limiting the bandwith. The curve (Fig. 8) shows a high
visibility over the whole scan range. However, in this case the photons have
approximately 70 nm FWHM and we have to assume some 100 ps spread. You may
feel more confident with this measurement, since only real detectors and no
black surfaces are involved. However, the moment of the collapse of the
wavefunction is better defined in a black surface. We can assume that the
lifetime of excited levels of molecules of the paint is very short, i.e.
that the absorbed photon energy is transformed to heat within less than 1
ps. In contrary, APD photon counters have a time jitter in the order of 300
ps, to some extent due to the fact that created electron-hole pair in the
absorbing takes more or less time to diffuse to the multiplying region. But
it's only there, where the irreversible proces from a quantum state to a
macroscopic state occurs, which is our definition of the collapse. So, in
fact we have an additional uncertainty in the timing of the collapse in the
order of 100 ps, the same order of magnitude as the dispersion induced
spread. The lower speed limit becomes then $\frac{1}{3}10^{6}c.$

One can assume that the collapse happens in some preferred frame, which is
not the frame of Geneva-Bellevue-Bernex. A reasonable candidate is the frame
of the cosmic background radiation. An analysis of our data shows that for
this frame a lower speed limit for the quantum information of 1.5x10$^{4}c$
can be given \cite{valerio, zbinden00}.

\section{Conclusions}

Entanglement is the main resource of Quantum Information Processing and is
at the core of the uneasiness many people face with the quantum world. It
thus deserves to be widely studied, both theoretically and experimentally.
In this work we have presented results from a first experiment in which both
the relativity of timing and entanglement of spatially separated systems are
central. Indeed, in the tested configuration the time ordering of the two
measurements of the quantum systems depend on the reference frame defined by
the two ''measurement apparatuses''. Each apparatus consists of an
interferometer with two outputs. One output is connected to a standard
photon counting detector, while the other output is connected to a ''passive
detector'', i.e. a detector which irreversibly absorbs the photon, but
spread the information in the environment without registering a signal in a
form readable for humans. We have argued that such ''passive detectors''
have the same physical effect on the photon and that the result of this
effect can be read of the active detector at the other output (if the photon
does not show up at one output, it is at the other output). Furthermore we
have argued that the crucial part of each measurement apparatus is the
detector which encounters the photon's wave-packet first. Hence, we arranged
the experiment such that the crucial parts of each measurement are the
''passive detectors'' and that these are in relative motion such that the
time ordering of the impacts of the two entangled photons on them depends on
the reference frame defined by these moving ''passive detectors''.

The results are always in accordance with QM, re-enforcing our confidence in
the possibility to base future understanding of our world and future
technology on quantum principles. To achieve our experiment we had to set
the 2-photon source very precisely at the center between the two
measurements, i.e. the two impacts on the detectors were simultaneous in the
Geneva reference frame to within 5 ps. This sets a lower bound on the speed
on quantum information to $10^7$c, i.e. seven orders of magnitude larger
than the speed of light.

The description of ''quantum measurements'' is notoriously difficult and
controversial. The results we have presented make it even more delicate to
give a realistic description with ''real collapses''. On the other side, the
reasoning we have followed opens new ways to test quantum mechanics and its
weird description of ''measurements''. The assumptions we made to achieve
this first experiment can and should be criticized. For instance, the
assumption that the detector is the crucial step in a measurement is at odds
with the idea that the collapse takes place in the reference frame
determines by this detector, as discussed in appendix B. But at least these
assumptions lead to a feasible experiment and will hopefully trigger new
proposals.

It is a great time for quantum physics. Both its foundations and its
potential applications are deeply explored by a growing community of
physicists, mathematicians, computer scientists and philosophers. We
explored experimentally some of the most counter intuitive predictions of
quantum theory, stressing the tension with relativity. Our results
contribute to the renewed interest for experimental challenges to the
interpretation of quantum mechanics and is relevant for the
realist-positivist debate. ''Experimental metaphysics'' questions \cite
{expmetaph} like ''what about the concept of states?'', ''the concept of
causalities?'' will have to be (re)considered taking into account the
results presented in this article. For example, our results make it more
difficult to view the ''projection postulate'' as a compact description of a
real physical phenomenon \cite{Pearle85,HPA89}.

\section*{Acknowledgments}

%=========================
This work would not have been possible without the financial support of the
''Fondation Odier de psycho-physique''. It also profited from support by
Swisscom and the Swiss National Science Foundation. \ We would like to thank
A. Suarez and V. Scarani for very stimulating discussions and H. Inamori for
preparing work during his stay in our lab.\newpage

\section*{Appendices}

\appendix

\section{Probabilities for moving observers}

%==============================
Let P and Q denote two projector acting on spatially separated Alice and Bob
systems, respectively. We shall use the identification $P\approx P\otimes %
\leavevmode\hbox{\small1\kern-3.8pt\normalsize1}$ and $\leavevmode%
\hbox{\small1\kern-3.8pt\normalsize1}\otimes Q\approx Q$. If the
measurements corresponding to P and Q are either before-after or
after-before, then the test-theory predicts the same probability as standard
QM, with $\psi \in {\cal H}_{1}\otimes {\cal H}_{2}$ the usual quantum
state: 
\begin{eqnarray}
&&Prob(++|b-a)=<Q>_{P\psi }<P>_{\psi } \\
&=&Prob(++|a-b)=<P>_{Q\psi }<Q>_{\psi } \\
&=&Prob(++|QM)=<P\otimes Q>_{\psi }
\end{eqnarray}
If, however, both measurements are before we postulate (inspired by Suarez
\& Scarani): 
\begin{equation}
Prob(++|b-b)=<P>_{\psi }<Q>_{\psi }\neq Prob(++|QM)
\end{equation}
The case after-after is the most delicate to guess. Inspired by Suarez'
intuition that in such a case each particle tries to guess what the other
would have done if it were before (as if the information from the other
particle would have got lost), we try the following postulate: 
\begin{eqnarray}
Prob(++|a-a) &=&<P\otimes Q>_{\psi }<P>_{Q\psi }<Q>_{P\psi }  \nonumber \\
&+<&P\otimes Q^{\perp }>_{\psi }<P>_{Q^{\perp }\psi }<Q>_{P\psi }  \nonumber
\\
&+<&P^{\perp }\otimes Q>_{\psi }<P>_{Q\psi }<Q>_{P^{\perp }\psi }  \nonumber
\\
&+<&P^{\perp }\otimes Q^{\perp }>_{\psi }<P>_{Q^{\perp }\psi }<Q>_{P^{\perp
}\psi }  \label{Ppp} \\
&\neq &Prob(++|QM)  \nonumber
\end{eqnarray}
where $<P>_{Q\psi }\equiv \frac{<Q\psi |P|Q\psi >}{<Q\psi |Q\psi >}=\frac{%
<P\otimes Q>_{\psi }}{<Q>_{\psi }}$. The idea is that Alice system evaluate
the projector P in either the state $Q\psi $ or $Q^{\perp }\psi $ depending
on its guess of Bob' system outcome. The situation is clearly symmetric.
Hence the 4 alternatives are weighted according to the standard outcome
probabilities (it does not matter whether it is Alice who guesses that Bob
was first, or whether it is Bob who assumes that Alice was first). In (\ref
{Ppp} each line corresponds to one possible guess, the first term giving the
corresponding guess probability (e.g. $<P\otimes Q>_{\psi }$ for both
guessing that theother had a positive outcom) and the two last terms the
corresponding outcomes probabilities (e.g. $<P>_{Q\psi }<Q>_{P\psi }$).

The other probabilities for the after-after configuration follow: e.g. $%
Prob(+-|a-a$ obtains from (\ref{Ppp}) by replacing $Q$ with $Q^\perp$.

A first consistency check is for product states: if $\psi=\alpha\otimes\beta$%
, then from (\ref{Ppp}) follows $Prob(++|a-a)=<P>_\alpha<Q>_\beta$.

As second consistency check let us compute: 
\begin{eqnarray}
Prob(++|a-a)&+&Prob(+-|a-a)=  \nonumber \\
&=& <P\otimes Q>_\psi (<P>_{Q\psi}<Q>_{P\psi}+<P>_{Q\psi}<Q^\perp>_{P\psi}) 
\nonumber \\
&+&<P\otimes Q^\perp>_\psi
(<P>_{Q^\perp\psi}<Q>_{P\psi}+<P>_{Q^\perp\psi}<Q^\perp>_{P\psi})  \nonumber
\\
&+&<P^\perp\otimes Q>_\psi
(<P>_{Q\psi}<Q>_{P^\perp\psi}+<P>_{Q\psi}<Q^\perp>_{P^\perp\psi})  \nonumber
\\
&+&<P^\perp\otimes Q^\perp>_\psi
(<P>_{Q^\perp\psi}<Q>_{P^\perp\psi}+<P>_{Q^\perp\psi}<Q^\perp>_{P^\perp\psi})
\\
&=& <P\otimes Q>_\psi <P>_{Q\psi} + <P\otimes Q^\perp>_\psi <P>_{Q^\perp\psi}
\nonumber \\
&+&<P^\perp\otimes Q>_\psi <P>_{Q\psi} + <P^\perp\otimes Q^\perp>_\psi
<P>_{Q^\perp\psi} \\
&=&<Q>_\psi <P>_{Q\psi} + <Q^\perp>_\psi <P>_{Q^\perp\psi} \\
&=&<P>_\psi
\end{eqnarray}
Accordingly, the local probabilities follow the standard laws, as they
should!

We compute now the correlation function 
\begin{eqnarray}
E(P,Q) &=&Prob(++)+Prob(--)-Prob(+-)-Prob(-+) \\
&=&<P\otimes Q>_{\psi }<\sigma _{P}>_{Q\psi }<\sigma _{Q}>_{P\psi } 
\nonumber \\
&+<&P\otimes Q^{\perp }>_{\psi }<\sigma _{P}>_{Q^{\perp }\psi }<\sigma
_{Q}>_{P\psi }  \nonumber \\
&+<&P^{\perp }\otimes Q>_{\psi }<\sigma _{P}>_{Q\psi }<\sigma
_{Q}>_{P^{\perp }\psi }  \nonumber \\
&+<&P^{\perp }\otimes Q^{\perp }>_{\psi }<\sigma _{P}>_{Q^{\perp }\psi
}<\sigma _{Q}>_{P^{\perp }\psi }  \label{Epq} \\
&&  \nonumber
\end{eqnarray}
where $\sigma _{P}\equiv P-P^{\perp }$ and $\sigma _{Q}\equiv Q-Q^{\perp }$.
Let us illustrate this conjecture for the singlet state with $\sigma _{P}=%
\vec{a}\vec{\sigma}$ and $\sigma _{Q}=\vec{b}\vec{\sigma}$: 
\begin{eqnarray}
E(\vec{a},\vec{b}) &=&\frac{1-\vec{a}\vec{b}}{4}\left( (\vec{a}\vec{b}%
)^{2}+(-\vec{a}\vec{b})^{2}\right)  \nonumber \\
&+&\frac{1+\vec{a}\vec{b}}{4}\left( (\vec{a}\vec{b})(-\vec{a}\vec{b})+(-\vec{%
a}\vec{b})(\vec{a}\vec{b})\right) \\
&=&-(\vec{a}\vec{b})^{3}  \label{Eab}
\end{eqnarray}
This may be more difficult to distinguish experimentally from the
before-before conjecture. Note that if $\vec{a}$ and $\vec{b}$ are parallel
(or anti-parallel), then the maximal anti-correlation (correlation) are
predicted. This is in accordance with the idea that each particle guesses
the other's output.

\section{Detectors as choice device and super-luminal communication}

%====================================================================
In this appendix we elaborate on the assumption that the collapse (i.e. the
outcome of the measurement) is determined by the detector. We show that
using this assumption one can device a situation in which quantum
nonlocality could be activated, that is use to signal at arbitrary high
speeds.

Let us return to figures 1. Why not leave the interferometers close to the
source and just put the absorbing surfaces at a distance. The detectors can
then also stay by the source, provided a long fiber spool is inserted to
ensure that the detections occur after the collapse (see Fig. 1c). So we are
looking at coincidences between two detectors side by side. When the wheel
is turning at Bob's the correlation should disappear. Since Bob can be very
far away from the source and the detection of photons must occur just
shortly after the potential arrival of the photons at Alice and Bob. Bob
could, by switching on and off the wheel, send signals back to the source at
superluminal speed.

The peaceful coexistence between quantum mechanics and relativity \cite
{Shimony83} is one of the most remarkable facts of physics! It is
notoriously difficult to modify quantum mechanics without activating quantum
non-locality, hence without breaking this peaceful co-existence \cite
{NGWeinberg,Czachor}. Weinberg has argued on this base that quantum
mechanics is part of the final theory \cite{Weinberg93}! In this appendix we
see once again that a proposed modification to basic quantum mechanics
requires also a radical modification of relativity. However, it is possible
that it is not the detector that triggers the collapse. The photons could
take the decision already at the beamsplitter and go out through one output
port, like in the Bohm-de-Bloglie pilot wave picture \cite{BellPilotWave}
(which much inspired Suarez). With the beam-splitter as choice-device
superluminal signaling is not possible (to our knowledge). A corresponding
experimental test would be more demanding, a beam-splitter would have to be
in motion. A clever way-out could be the use of an acousto-optical modulator
representing a beam-splitter moving with the speed of the acoustic wave. We
are working on such an experiment.

\section{Figures}

\begin{figure}
\infig{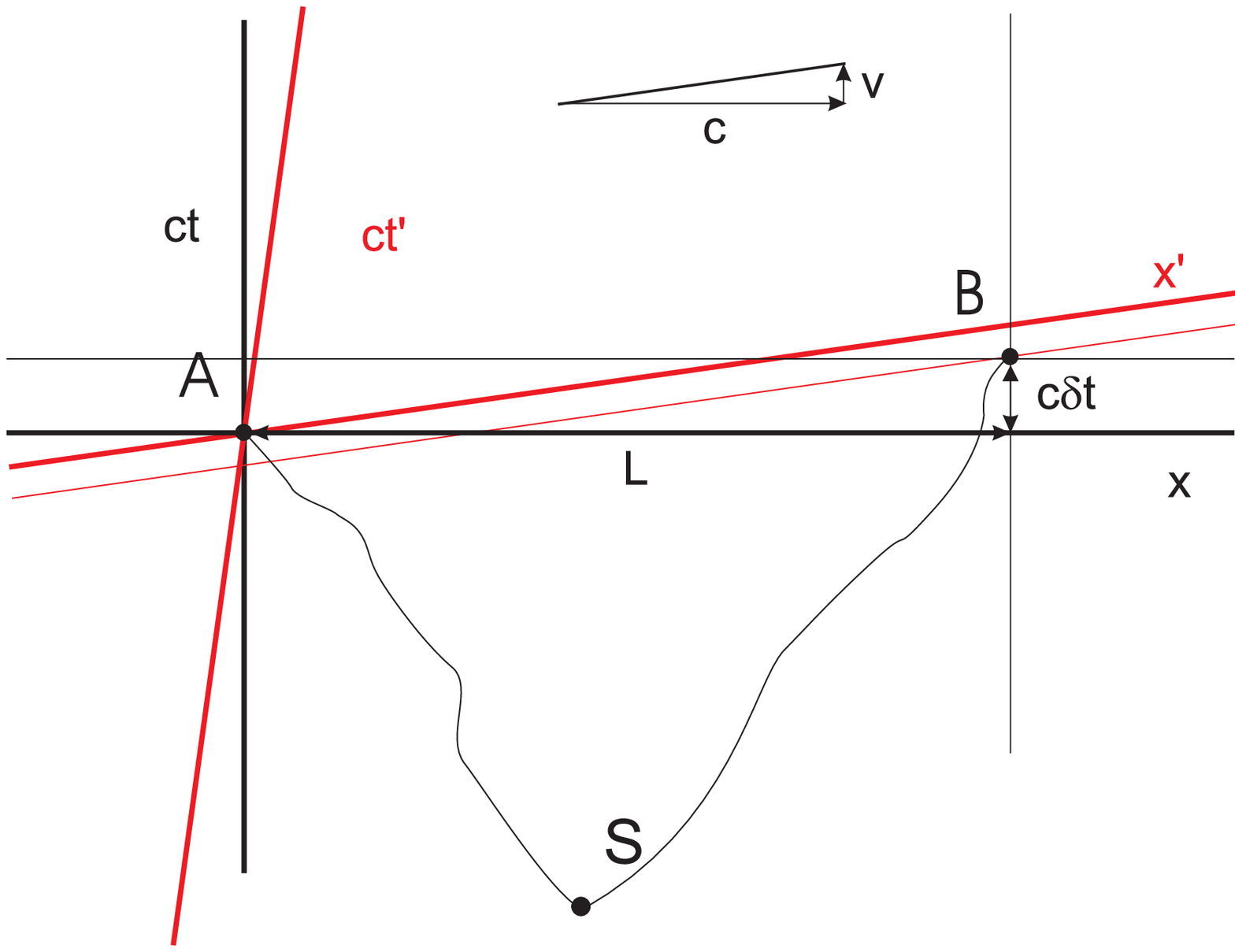}{0.80\columnwidth}
\caption{Space-time diagram of a Bell experiment with moving observer. In the
referential frame of the moving observer (x',ct') event B happens before
event A. Note that the source does not have to be located in the middle
between the two detectors. Only the optical path lengths must be the same.
}
\label{fig1} 
\end{figure}

\begin{figure}
\infig{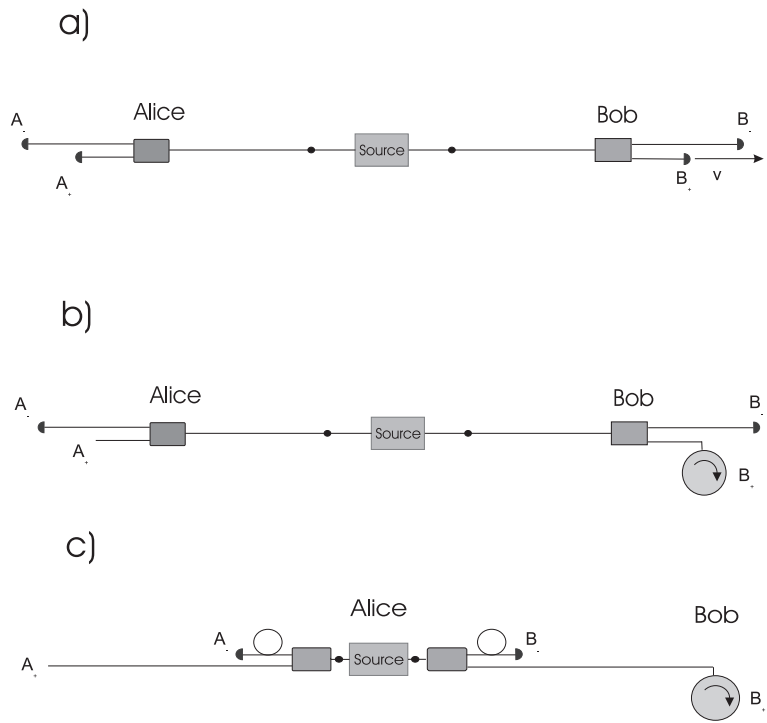}{0.80\columnwidth}
\caption{Schematic of the experimental arrangement with a moving detector (a).
The linearly moving detector can be replaced by a rotating absorber (b). By
placing the detectors close to the source (c), a breakdown of the nonlocal
correlations could be exploited for superluminal signalling (see Appendix B).}
\label{fig2} 
\end{figure}

\begin{figure}
\infig{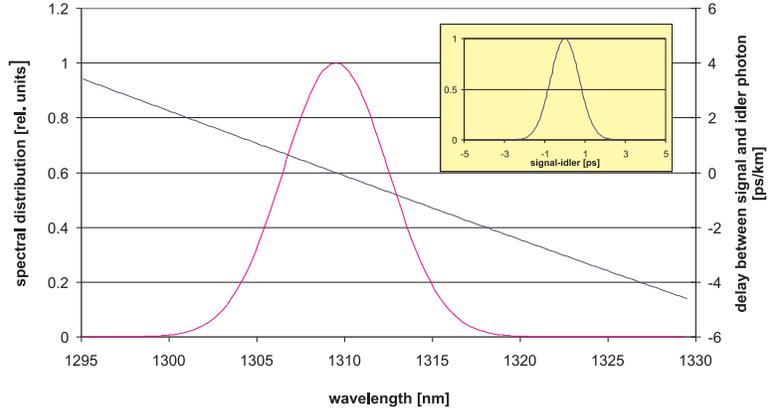}{0.80\columnwidth}
\caption{Spectral distribution of photon pairs (10nm FWHM interference filter
centered at 1309.5nm ) and differential group delay for a fibre with $%
\lambda _{0}=1310nm$. Since the photons are centered close to $\lambda _{0},$
the most of dispersion induced delay between signal and idler photon is
cancelled. The inset gives the distribution of time delay between idler and
signal after 1 km of fibre.}
\label{fig3} 
\end{figure}

\begin{figure}
\infig{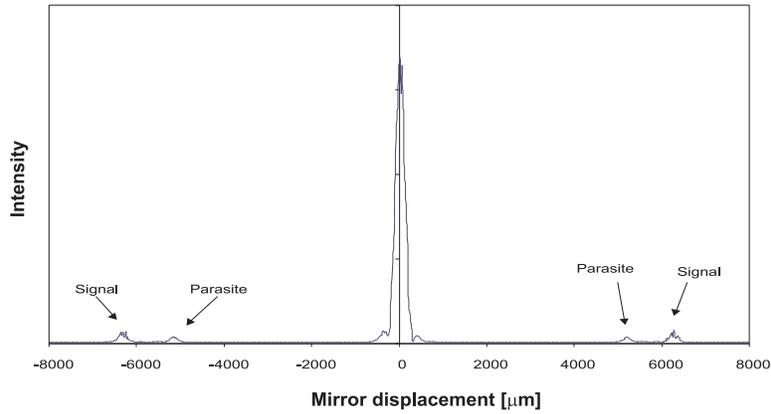}{0.80\columnwidth}
\caption{Typical scan of the low coherence interferometer. The signal is low
due to the losses after 20 km of fibre (roundtrip). Nevertheless the
position of peak can be determined with a precision of about 100$\mu m$.
There is a parasite signal due to a reflection close to the source.}
\label{fig4} 
\end{figure}

\begin{figure}
\infig{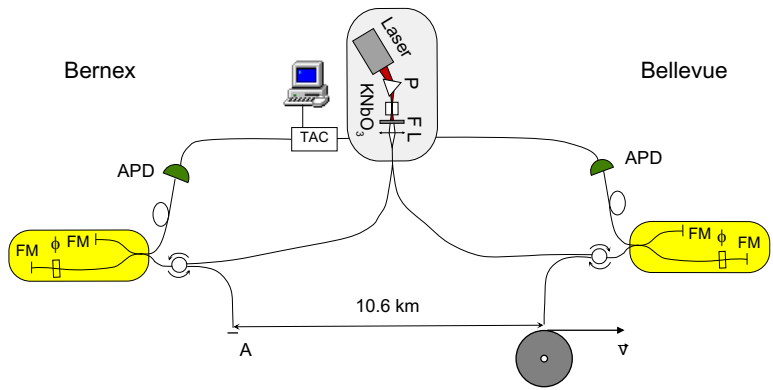}{0.80\columnwidth}
\caption{Schematic of the experiment.}
\label{fig5} 
\end{figure}

\begin{figure}
\infig{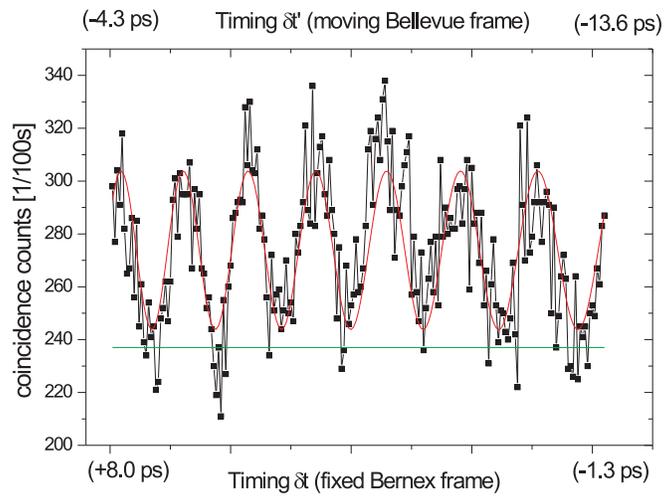}{0.80\columnwidth}
\caption{2-photon interference fringes measured over 6 h while the
optical distance to Bernex was slowly overpassing that to Bellevue. Positive
time values indicate that the detection occued first in Bernex. According
the moving reference frame, dectection occured first in Bellvue over the
whole scan range. Despite this different chronology of the events no change
in the visibility can be observed.}
\label{fig6} 
\end{figure}

\begin{figure}
\infig{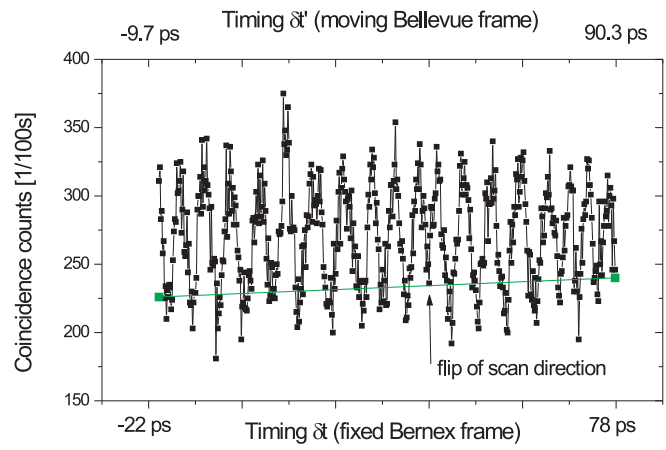}{0.80\columnwidth}
\caption{2-photon interference fringes measured over 14 h while the
optical distance to Bellevue was slowly overpassing that to Bernex. At some
time we have negative values in fixed Bernex frame, indicating that the
detection occurs first in Bellevue, and positive values in the moving
Bellvue frame, indicating that detection detection occurs first in Bernex.
Despite this after-after constellation no reduced visibility can be seen.}
\label{fig7} 
\end{figure}

\begin{figure}
\infig{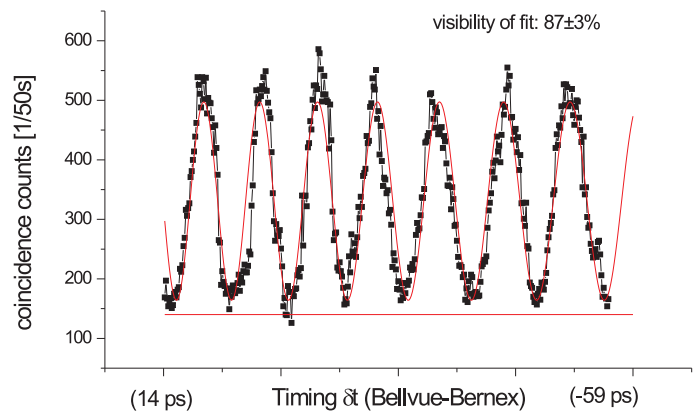}{0.80\columnwidth}
\caption{2-photon interference fringes measured with two aligned detectors
over 5h30 while the optical distance to Bernex was slowly overpassing that
to Bellevue. At a certain time the two fixed detectors are at exactly the
same distance from source.}
\label{fig8} 
\end{figure}

\end{document}